\documentclass[physrev,twocolumn,superscriptaddress,floatfix,altaffillsymbol]{revtex4-1}

\usepackage{graphicx} % for graphics
\usepackage{bm} % bold math
\usepackage{amssymb} % math stuff
\usepackage{amsmath} % math stuff
\usepackage{color} % color text
\usepackage{subfigure} % divided figures
\usepackage{textcomp} % symbols
\usepackage[bookmarks=true]{hyperref} % links and page setup
\usepackage[utf8]{inputenc}

\usepackage{amsfonts}
\usepackage{multirow}
\usepackage{longtable, tabu}
\usepackage{tablefootnote}
\usepackage{tabularx}
\usepackage{booktabs}
\hypersetup{colorlinks=true,pdfborder={0 0 0},citecolor=blue,linkcolor=blue,urlcolor=blue}

\begin{document}
%\raggedbottom
\title{Collective 2p-2h intruder states in $^{118}$Sn studied via $\beta$-decay of $^{118}$In using GRIFFIN}

\date{\today}

\author{K.~Ortner}
\email{kortner@sfu.ca}

\affiliation{Department of Chemistry, Simon Fraser University, Burnaby, British Columbia V5A 1S6, Canada}

\author{C.~Andreoiu}

\affiliation{Department of Chemistry, Simon Fraser University, Burnaby, British Columbia V5A 1S6, Canada}

\author{M. Spieker}
%\email[]{mspieker@fsu.edu}
\altaffiliation{Present address:Department of Physics, Florida State University, Tallahassee, FL 32306, USA}
\affiliation{National Superconducting Cyclotron Laboratory, Michigan State University, 640 South Shaw Lane, East Lansing, MI 48824, USA}

%\author{M. Spieker}
%
%\altaffiliation{Present address: Department of Physics, Florida State University, 77 Chieftan Way, Tallahassee, FL 32306-4350}
%
%\affiliation{NSCL, 640 South Shaw Lane, East Lansing, MI 48824, USA}

\author{G.C.~Ball}

\affiliation{TRIUMF, 4004 Wesbrook Mall, Vancouver, British Columbia V6T 2A3, Canada}

\author{N.~Bernier}

\affiliation{TRIUMF, 4004 Wesbrook Mall, Vancouver, British Columbia V6T 2A3, Canada}

\affiliation{Department of Physics and Astronomy, University of British Columbia, Vancouver, British Columbia V6T 1Z4, Canada}

\author{H.~Bidaman}

\affiliation{Department of Physics, University of Guelph, Guelph, Ontario N1G 2W1, Canada}

\author{V.~Bildstein}

\affiliation{Department of Physics, University of Guelph, Guelph, Ontario N1G 2W1, Canada}

\author{M.~Bowry}

\altaffiliation{Present address: School of Computing, Engineering and Physical Sciences, University of the West of Scotland, Paisley PA1 2BE, United Kingdom}

\affiliation{TRIUMF, 4004 Wesbrook Mall, Vancouver, British Columbia V6T 2A3, Canada}

\author{D.S.~Cross}

\affiliation{Department of Chemistry, Simon Fraser University, Burnaby, British Columbia V5A 1S6, Canada}

\author{M.R.~Dunlop}

\affiliation{Department of Physics, University of Guelph, Guelph, Ontario N1G 2W1, Canada}

\author{R.~Dunlop}

\affiliation{Department of Physics, University of Guelph, Guelph, Ontario N1G 2W1, Canada}

\author{F.H.~Garcia}

\affiliation{Department of Chemistry, Simon Fraser University, Burnaby, British Columbia V5A 1S6, Canada}

\author{A.B.~Garnsworthy}

\affiliation{TRIUMF, 4004 Wesbrook Mall, Vancouver, British Columbia V6T 2A3, Canada}

\author{P.E.~Garrett}

\affiliation{Department of Physics, University of Guelph, Guelph, Ontario N1G 2W1, Canada}

\author{J.~Henderson}

\altaffiliation{Present address: Lawrence Livermore National Laboratory, Livermore, California 94550, USA}

\affiliation{TRIUMF, 4004 Wesbrook Mall, Vancouver, British Columbia V6T 2A3, Canada}

\author{J.~Measures}

\affiliation{TRIUMF, 4004 Wesbrook Mall, Vancouver, British Columbia V6T 2A3, Canada}

\affiliation{Department of Physics, University of Surrey, Guildford GU2 7XH, United Kingdom}

\author{B.~Olaizola}

\affiliation{TRIUMF, 4004 Wesbrook Mall, Vancouver, British Columbia V6T 2A3, Canada}

\author{J.~Park}

\altaffiliation{Present address: Department of Physics, Lund University, 22100 Lund, Sweden}

\affiliation{TRIUMF, 4004 Wesbrook Mall, Vancouver, British Columbia V6T 2A3, Canada}

\affiliation{Department of Physics and Astronomy, University of British Columbia, Vancouver, British Columbia V6T 1Z4, Canada}

\author{C.M.~Petrache}

\affiliation{Université Paris-Saclay, CNRS/IN2P3, IJCLab, 91405, Orsay, France}

\author{J.L.~Pore}

\altaffiliation{Present address: Lawrence Berkeley National Laboratory, Berkeley, California 94720, USA}

\affiliation{Department of Chemistry, Simon Fraser University, Burnaby, British Columbia V5A 1S6, Canada}

\author{K.~Raymond}

\affiliation{Department of Chemistry, Simon Fraser University, Burnaby, British Columbia V5A 1S6, Canada}

\author{J.K.~Smith}

\altaffiliation{Present Address: Department of Physics, Pierce College, Puyallup, Washington 98374, USA}

\affiliation{TRIUMF, 4004 Wesbrook Mall, Vancouver, British Columbia V6T 2A3, Canada}

\author{D.~Southall}

\altaffiliation{Present address: Department of Physics, University of Chicago, Chicago, IL 60637, USA}

\affiliation{TRIUMF, 4004 Wesbrook Mall, Vancouver, British Columbia V6T 2A3, Canada}

\author{C.E.~Svensson}

\affiliation{Department of Physics, University of Guelph, Guelph, Ontario N1G 2W1, Canada}

\author{M.~Ticu}

\affiliation{Department of Chemistry, Simon Fraser University, Burnaby, British Columbia V5A 1S6, Canada}

\author{J.~Turko}

\affiliation{Department of Physics, University of Guelph, Guelph, Ontario N1G 2W1, Canada}

\author{K.~Whitmore}

\affiliation{Department of Chemistry, Simon Fraser University, Burnaby, British Columbia V5A 1S6, Canada}

\author{T.~Zidar}

\affiliation{Department of Physics, University of Guelph, Guelph, Ontario N1G 2W1, Canada}

\begin{abstract}

The low-lying structure of semi-magic $^{118}$Sn has been investigated through the $\beta$-decay of $^{118}$In ($T_{1/2}=4.45$~min) to study shape coexistence via the reduced transition probabilities of states in the 2p-2h proton intruder band. This high-statistics study was carried out at TRIUMF-ISAC with the GRIFFIN spectrometer. In total, 99 transitions have been placed in the level scheme with 43 being newly observed. Three low-lying $\gamma$-ray transitions with energies near 285~keV have been resolved from which the 2$^+_{\mathrm{intr.}} \rightarrow 0^+_{\mathrm{intr.}}$ 284.52-keV transition was determined to have half of the previous branching fraction leading to a $B(E2;2^+_2\rightarrow 0^+_2)$ of 21(4) W.u. compared to 39(7) W.u. from the previous measurement. Calculations using $sd$ IBM-2 with mixing have also been made to compare the experimental $B(E2)$ values to the theoretical values and to make comparisons to the $^{114,116}$Sn isotopes previously studied using the same theoretical model.

\end{abstract}

\maketitle

\section{Introduction}

%isotopic chain minimum around 116 observations
The semi-magic isotopes of Sn continue to be of great interest. They are benchmark nuclei for state-of-the-art shell-model calculations and offer a strong foundation to our understanding of shape evolution in the $Z=50$ region (see, e.g., Ref.~\cite{togashi2018}). While the Sn isotopes are close to spherical in their ground state, deformed bands built on excited 0$^{+}$ states are observed throughout this isotopic chain and have been interpreted as having two-particle-two-hole, 2p-2h, character~\cite{bron,heyde2011}. The presence of these deformed intruder states is considered to be an important feature in the $Z=50$ region~\cite{garrett2016,spieker2018,heyde2011} and the degree of mixing between deformed and normal states needs to be further explored~\cite{heyde2011}.

Enhanced cross sections of the excited 0$^+$ states in even-even $^{108-118}$Sn were identified in the ($^3$He,$n$) experiments by Fielding \textit{et al.}~\cite{fielding}, and it was later suggested that the observed 0$^+_2$ state in even-even $^{112-118}$Sn is the bandhead of an intruding rotational band due to 2p-2h proton excitations across the shell gap by Br\"on \textit{et al.}~\cite{bron}. In a recent $\beta^-$ decay study of $^{116}$In to $^{116}$Sn, a newly obtained $B$($E$2; $2^+_2 \rightarrow 0^+_3$) value suggested the 0$^+_3$ state as the 2p-2h bandhead instead of the 0$^+_2$~\cite{pore116}. Furthermore, $sd$ IBM-2 with mixing calculations were made to test the mixing between the intruder and normal configurations in $^{116}$Sn~\cite{petrache} and $^{114}$Sn~\cite{spieker2018}. In both cases the conclusion was that the 0$^+_3$ is the intruding bandhead but is strongly mixed with the 0$^+_2$ state. 

Although the intruder band lies at lower excitation energy in $^{118}$Sn---the 2$^+_{\mathrm{intr.}}$ state is 14~keV below the 0$^+_3$ level---obtaining accurate $B$($E$2) values can indicate the degree of collectivity of the intruder band and the amount of mixing between deformed and normal states~\cite{bohrmott}.

Many studies have been performed on $^{118}$Sn providing a comprehensive level scheme which includes the 2p-2h intruder band built on the excited 0$_2^{+}$ state at 1758~keV~\cite{nndc}. 
Two studies, $\beta$-decay of $^{118}$In~\cite{raman} and $(n,n'\gamma)$~\cite{mikhailov}, contributed most of the low-lying observables to the $^{118}$Sn level scheme. One notable discrepancy between these studies, which needs clarification, is related to the $I_{\gamma}(2^+_3 \rightarrow 2^+_2)$ for 285.3~keV and the $I_{\gamma}(2^+_2 \rightarrow 0^+_2)$ for 284.6~keV~\cite{mikhailov,raman}. In each of these experiments, both the 2$^+_2$ intruder state and the 2$^+_3$ state were populated. However, in the ($n,n'\gamma$) study, only the 284.6-keV transition was observed~\cite{mikhailov}, while the $^{118}$In $\beta$-decay study only observed the 285.3-keV transition~\cite{raman}. It is likely that the intensities of these transitions have been grouped together in the ENSDF~\cite{nndc} and as a result, the adopted $B$($E$2; 2$_2^{+} \rightarrow 0_2^{+}$) of 39(7) W.u. is too large. Since the characterization of the collectivity in the normal and intruder configurations is important, the present work sought to resolve this issue in $^{118}$Sn.

A high-statistics experiment to investigate the decay properties of the proton 2p-2h band in $^{118}$Sn using the $\beta$-decay of $^{118}$In and the high-resolution GRIFFIN spectrometer located at TRIUMF-ISAC~\cite{griffin} has been performed. New results, which include updated $B(E2)$ values, and a discussion of the new observations are presented. Furthermore, $sd$ IBM-2 calculations with mixing have been performed to further elucidate the character of the states of interest. 

%concluding comment about results and B(E2)

\section{Experiment}

Measurements of the $\beta$-decay of $^{118}$In were performed at the TRIUMF Isotope Separator and ACcelerator (ISAC) facility which houses the Gamma Ray Infrastructure For Fundamental Investigations of Nuclei (GRIFFIN)~\cite{griffin}. GRIFFIN is a high-efficiency $\gamma$-ray spectrometer consisting of 16 high-purity germanium (HPGe) clover detectors~\cite{hpge} which was coupled to the ancillary SCintillating Electron Positron Tagging ARray (SCEPTAR), comprised of 20 plastic scintillators for tagging $\beta$-particles~\cite{griffin} and was fixed within a 20mm Delrin shield. The detector signals were read out and processed by the GRIFFIN data acquisition system~\cite{daq}.

A radioactive beam of $^{118}$In was mass separated from the reaction products of a 9.8~\textmu A, 480~MeV proton beam impinged onto a UC$_x$ target. A high-purity beam was obtained using the Ion-Guide Laser Ion Source (IG-LIS) to suppress isobaric contaminants. The beam was transported and implanted into Mylar tape at the center of the GRIFFIN chamber. The isotope $^{118}$In decays via the $1^+$ ground state with $T_{1/2}$ = 5~s~\cite{hl1}, a $5^+$ isomeric state with $T_{1/2}$ = 4.45~min~\cite{amaral} and an $8^-$ isomeric state with $T_{1/2}$ = 8.5~s~\cite{hl8}. The tape was cycled after five minutes of implant and five minutes of decay to obtain statistics which favored observation of the 5$^+$ isomeric state, which populates the states of interest in $^{118}$Sn. Analysis was performed on data after 25~s of decay, reducing the contributions from the 1$^+$ and 8$^-$ by five and three halflives, respectively. Furthermore, nearly 99~\% of the 8$^-$ state decays internally to the 5$^+$ state through a 138.5~keV $\gamma$ ray, and 95~\% of the 1$^+$ state $\beta$-decays to the 0$^+$ ground state in $^{118}$Sn. The tape was moved out of the chamber and into a lead box after each decay cycle to start a new implant cycle. The total run time was approximately 80 minutes during which $2\times10^9$ $\gamma$-singles events and $1\times10^9$ $\gamma$-$\gamma$\ coincidence events were recorded. Since the isotope of interest was sufficiently free of isobaric contaminants and the implant rate was quite high, SCEPTAR was not used to generate $\beta$-$\gamma$ coincident spectra. 

The relative efficiency of the 16 HPGe clovers in addback mode~\cite{griffin} was determined using standard sources of $^{56}$Co, $^{60}$Co, $^{133}$Ba and $^{152}$Eu. Addback is a technique in which coincident $\gamma$-ray energies between adjacent crystals in a single clover detector are summed. Compton scattered $\gamma$-ray events are recovered increasing the peak-to-total ratio, and ultimately increasing the total photopeak efficiency. For this analysis, the addback mode was applied to determine the peak intensities. The peak centroids and areas were obtained using maximum-likelihood fitting of a modified Gaussian with parameters for skewedness, linear or quadratic background and step sizes. Corrections to the $\gamma$-ray intensities due to summing were determined using a matrix of $\gamma$-$\gamma$\ coincidences between detector pairs separated by 180$^{\circ}$ as described in Ref.~\cite{griffin}. 

Energy calibrations were made using a linear fit between two strong photopeaks and then non-linearity corrections were made using many of the well known photopeaks in $^{118}$Sn. A systematic uncertainty of 0.2~keV was determined on the $\gamma$-ray energies based on the non-linearity residuals applied to the calibration sources. Cross-talk corrections were made using Compton events in the $\gamma$-$\gamma$ coincidence matrix from the $^{60}$Co source as outlined in section 4.1.1 in Ref.~\cite{griffin}. The uncertainty on $\gamma$-ray intensities is based on the uncertainty in peak areas, background estimations, relative efficiencies, and summing corrections. 

For weak transitions, $\gamma$-$\gamma$\ coincidence matrices were used to obtain their energies and intensities. This used a method of gating on a stronger $\gamma$-ray transition directly below the weak $\gamma$-ray in the same cascade~\cite{gating, coincidence}. To correct for summing in gated coincidence spectra, coincidence matrices of detector pairs separated by 180$^{\circ}$ were constructed from energy gates corresponding to the same gates used in the $\gamma$-$\gamma$\ coincidence matrices. The $\gamma$-$\gamma$ coincidences between any two of the 64 GRIFFIN crystals were also used to obtain $\gamma$-$\gamma$ angular correlations to determine the multipole mixing ratios $\delta$ (see Section~\ref{sec_angcorr}). 

To calculate the log($ft$) values, the $\beta$-feeding to individual levels needs to be determined. This was done through an intensity balance of $\gamma$-intensity depopulating a level minus $\gamma$-intensity populating the same level. To obtain absolute $\beta$-feeding of each level, the total ground-state feeding was used to normalize the relative $\beta$-feeding.

The $B(E2)$ values for transitions with known lifetime measurements, mixing ratios $\delta$ (from~\cite{nndc} or our measured values as described in Section~\ref{sec_angcorr}), or for transitions which are assumed pure $E2$, were calculated using our measured $\gamma$-ray branching ratios $BR_{\gamma}$.  All $BR_{\gamma}$ used in these calculations, and in the log$ft$ calculations took into account internal conversion. The internal conversion coefficients were applied to $\gamma$-ray intensities for energies below 600~keV and the coefficients were taken from~\cite{nndc} when possible, or calculated using BrIcc~\cite{bricc}. 

\section{Results}
\label{results}
\subsection{Level Scheme}

\begin{figure*}

     \includegraphics[width=\textwidth]{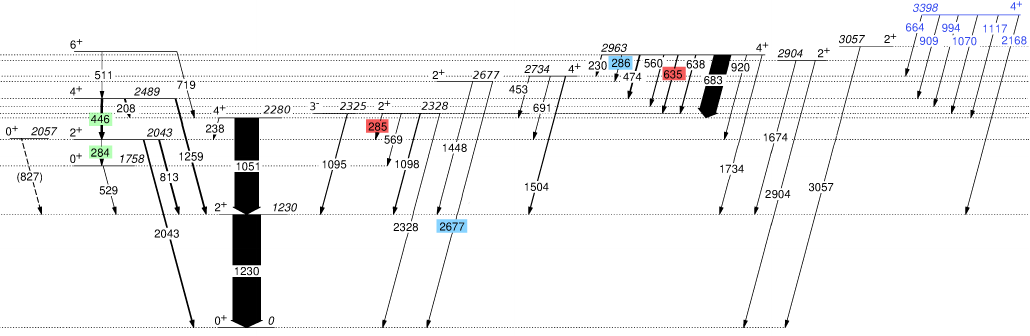}
		\caption{Partial level scheme of $^{118}$Sn populated in the $\beta$-decay of the 5$^+$ isomer of $^{118}$In. The widths of the arrows represent the relative intensities of the $\gamma$-rays. Included, but not observed, is the $0^+_3$ level at 2057~keV. 
		The 3398-keV level (blue) is newly placed based on five newly observed transitions depopulating it. 
		} 
\label{partial}
\end{figure*}
%round level energies
%Figure at start of sentence Fig. in sentence
In this $\beta$-decay study of the 5$^+$ isomer of $^{118}$In, 99 $\gamma$ rays have been assigned to 23 low-lying excited states in $^{118}$Sn. Of these transitions, 43 were newly observed, including 6 which depopulate a newly placed level at 3398~keV. Table~\ref{levels} shows all $\gamma$ rays observed in this experiment from an initial level of spin and parity \textit{J$^\pi_i$} to a final level of spin and parity \textit{J$^\pi_f$}, and their corresponding intensities, branching ratios, and $B(E2)$ transition strengths, and compares them to the ENSDF~\cite{nndc}. 

Of the 51 $\gamma$ rays placed by Raman \textit{et al.}~\cite{raman}, 49 were observed with relative intensities for most of the strong transitions in reasonably good agreement. However, there are several transitions where the $\gamma$-ray intensities and/or branching ratios differ by more than 2 sigma. Some of these can be easily explained from the fact we used $\gamma$-$\gamma$ coincidence matrices, while Raman states they did not. 

A 1265.14-keV transition from the 3753-keV level was observed by Raman in a singles spectra with an intensity of 0.138(9). We observe the 1265.14-keV transition and a newly observed 1264.34-keV transition from the 3592-keV level with intensities of 0.064(6) and 0.051(6), respectively. It is suspected that these were not resolved as two distinct transitions and the intensity placed by Raman is the sum of their intensities.

Raman identified the transition of 971.0 keV from the 3374-keV level and assigned an intensity of 0.35(6) based on an intensity balance of the 2403-keV level which the 971.0-keV transition populates. The remaining intensity of 0.32(7) was assigned to the 971.6-keV transition from the 3460-keV level. We observed more intensity feeding into the 2403-keV level, indicating that the 971.0-keV transition should have less intensity from what Raman observed. Furthermore, a third transition of 970.8 keV was resolved with an intensity of 0.110(5) from the 3704-keV level. Our summed intensity of 971.0-keV, 971.6-keV and 970.8-keV transitions is in agreement with the summed intensities Raman observed, although we determined an intensity of only 0.139(6) for the 971.0-keV transition compared to 0.35(6). 

Another discrepancy is the intensity given to the 1094.10-keV transition from the 3374-keV level. We observe, from $\gamma$-$\gamma$ coincidence, an intensity of 0.548(23) compared to 0.805(20). It is unclear how Raman was able to separate this from the 1094.98-keV transition depopulating the 2327-keV level without $\gamma$-$\gamma$ coincidence. 

One interesting difference is between our $B(E2)$ value of $<$42 W.u. for the 1098.2-keV transition depopulating the 2327-keV level compared to $B(E2)$ value of $<$20 W.u. listed in the ENSDF~\cite{nndc}. It is unclear how the value of $<$20~W.u. was established given our measured absolute $BR_\gamma$ for the 1098.2-keV transition is nearly the same as listed in the ENSDF\, \cite{nndc}.

We did not observe the 756-keV and the 1116-keV transitions assigned by Raman to the decay of a new level in $^{118}$Sn at 3159 keV. However, a 1117.3-keV transition was observed and originates from a newly observed 3398-keV level as shown in Figure~\ref{partial}. All of the $\gamma$ rays which depopulate this level were observed through $\gamma$-$\gamma$ coincidences and a spin of $J^\pi = 4^+$ was assigned to this level based on the log$ft$ value of 6.72 (shown in Table~\ref{logft}) and on the observation of 994.18-keV, 1070.06-keV and 2168.3-keV transitions which feed into the $2^+_4$, $2^+_3$ and $2^+_1$ levels, respectively. 

New transitions of 2524.3~keV and 1711.2~keV were observed and assigned to the decay of the 3754-keV level to the 1230-keV 2$^+_1$ and 2043-keV 2$^+_2$ levels, respectively. This suggests a spin assignment of 4$^+$ for the 3754-keV level which was previously given a \textit{J} = 4, 5, or 6 with no parity~\cite{raman}. The 4$^+$ assignment is also reasonable given the log$ft$ value of 5.874(14), as shown in Table~\ref{logft}, which is consistent with an allowed transition of $\Delta J = 1$ and no change in parity.

Spin assignments of $4^+$ were also made for the 3816.19-keV and 3838.33-keV levels based on observed transitions to 2$^+$ states and on the log$ft$ values in Table~\ref{logft}. The previous assignment of $J = 1^+$, 2$^+$, or 3$^+$ to the 3816.19-keV level would indicate at least a second forbidden transition which is not in line with the log$ft$ value obtained. Similarly, the 3838.33-keV level was previously assigned a $J = 4$  with no definite parity. This should be a $4^+$ based on the log$ft$ value.

A 1734-keV transition from the 2963-keV level had been observed previously by Hattula \textit{et al.}~\cite{hattula} and do Amaral \textit{et al.}~\cite{amaral}, but no evidence for this transition was reported by Raman \textit{et al.}~\cite{raman}. A significant amount of summing of the 683-keV transition with the 1051-keV transition gives rise to uncertainty for a 1734-keV transition which can be understood from Figure~\ref{levels}. However, summing corrections were performed for the 1734-keV photopeak and its relative intensity was determined to be 0.446(13), in agreement with the earlier reports~\cite{hattula,amaral}.

Non-zero $\beta$-feeding intensity is observed to the 2$^+$ 2904-keV and 3057-keV levels corresponding to unique second forbidden transitions (see Table~\ref{logft}). The log$ft$ values are 10.08(6) and 10.84(22), respectively, which are lower than expected. Higher than expected $\beta$ intensity is likely due to unobserved transitions. For instance, it is possible that there is a 59.5-keV transition connecting the 4$^+$ 2963-keV level to the 2904-keV level. However, this transition, if present, was not observed due to the low efficiency for $\gamma$ ray detection at this energy and the high probability for internal conversion.

\LTcapwidth=\textwidth
\begin{longtable*}{@{\extracolsep{\fill}}l l l l l l l | l l l}
\caption[table]{Levels in $^{118}$Sn populated by the $\beta^-$ decay of the $5^+$ isomer of $^{118}$In ($E_x = 60$~keV, $T_{1/2} = 4.45$~min). The relative intensity of the observed transitions, $I_{\gamma}$, are compared to the previous $\beta^-$ decay study~\cite{raman}, and the branching ratios, $BR_\gamma$, and $B(E2)$ values are compared to the ENSDF~\cite{nndc}. }\\
\hline
\hline
\toprule
\midrule
$E_{level}$ & $T_{1/2}$ (ps) & $J^{\pi}_i \rightarrow J^{\pi}_f$ & $E_{\gamma}$ & $I_{\gamma,rel}$ & $BR_{\gamma,rel}$ & $B(E2)$ & $I_{\gamma,rel}$ & $BR_{\gamma,rel}$ & $B(E2)$ \\
$[\text{keV}]$ & Ref.~\cite{nndc} & & $[\text{keV}]$ &  &  & $[\text{W.u.}]$ & Ref.~\cite{raman} & Ref.~\cite{nndc} & $[\text{W.u.}]$~\cite{nndc}\\
\midrule
\endfirsthead
\midrule
\endfoot
\caption*{Table I.$(Continued)$.}\\
\toprule
\midrule
$E_{level}$ & $T_{1/2}$ (ps) & $J^{\pi}_i \rightarrow J^{\pi}_f$ & $E_{\gamma}$ & $I_{\gamma}$ & $BR_\gamma$ & $B(E2)$ & $I_{\gamma}$ & $BR_\gamma$ & $B(E2)$\\
$[\text{keV}]$ & ref.~\cite{nndc} & & $[\text{keV}]$ &  &  & $[\text{W.u.}]$ & ref.~\cite{raman} & ref.~\cite{nndc} & $[\text{W.u.}]$~\cite{nndc}\\
\midrule
\endhead
\endlastfoot
\hline
1229.50(10)	&	0.49(2) 	&	$2^+_1 \rightarrow 0^+_1$	&	1229.57(20)	&	100			&	100	&	12.1(5)	&	100	&	100	&	12.1(5)	\\
1758.24(14)	&	21(3) 		&	$0^+_2 \rightarrow 2^+_1$	&	528.70(20)	&	0.129(4)	&	100	&	19(3)	&		&	100	&	19(3)	\\
2042.62(10)	&	2.9(4) 		&	$2^+_2 \rightarrow 0^+_2$	&	284.52(20)	&	0.051(7)	&	1.31(17)	&	21(4)	&		&	2.5(2)	&	39(7)	\\
			&				&	$2^+_2 \rightarrow 2^+_1$	&	813.11(21)	&	3.91(9)		&	100.0(23)	&	7.2(10)	&	3.88(12)	&	100.0(24)	&	6.9(1)	\\
			&				&	$2^+_2 \rightarrow 0^+_1$	&	2042.70(22)	&	3.27(10)	&	83.6(26)	&	0.072(10)	&	3.63(8)	&	92.2(25)	&	0.075(11)\\
2280.21(11)	&	0.76(13) 	&	$4^+_1 \rightarrow 2^+_2$	&	237.80(22)	&	0.050(4)	&	0.058(5)	&	16(3)	&	0.04(2)	&	0.05(2)	&	14(7)	\\
			&				&	$4^+_1 \rightarrow 2^+_1$	&	1050.54(20)	&	85(2)		&	100.0(26)	&	17(3)	&	84.4(26)	&	100(3)	&	17(3)	\\
2324.29(21)	&	0.19$^{+0.04}_{-0.03}$ 	&	$3^-_1 \rightarrow 2^+_1$&	1094.98(63)&1.46(7)	&	100(5)			&		&	1.5(5)	&	100(4)	&		\\
			&				&	$3^-_1 \rightarrow 0^+_1$	&	2323.9(3)	&	0.0148(13)	&	1.02(9)	&		&	&	1.1(1)	&		\\
2327.73(12)	&	$>$0.2 		&	$2^+_3 \rightarrow 2^+_2$	&	285.26(22)	&	0.038(14)	&	2.3(8)	&	&	0.081(10)	&	5.1(6)	&		\\
			&				&	$2^+_3 \rightarrow 0^+_2$	&	569.39(20)	&	0.041(2)	&	2.40(14)	&	$<$ 26	&		&		&		\\
			&				&	$2^+_3 \rightarrow 2^+_1$	&	1098.2(6)	&	1.70(7)		&	100(4)	&	$<$ 42	&	1.6(3)	&	100(19)	&	$<$20	\\
			&				&	$2^+_3 \rightarrow 0^+_1$	&	2327.7(6)	&	0.326(13)	&	19.1(8)	&	$<$ 0.19	&	0.374(12)	&	23.4(8)	&		\\
2403.05(11)	&	0.18$^{+0.08}_{-0.04}$ 	&	$2^+_4 \rightarrow 2^+_2$	&	360.67(22)		&	0.0121(18)	&	0.91(13)	&		&		&	1.8(2)	&		\\
			&				&	$2^+_4 \rightarrow 0^+_2$	&	644.73(20)	&	0.0190(8)	&	1.44(6)	&	12(5)	&		&		&		\\
			&				&	$2^+_4 \rightarrow 2^+_1$	&	1173.44(22)	&	1.32(3)		&	100.0(26)	&	17(7)	&	1.43(5)	&	100(3)	&	22(10)	\\
			&				&	$2^+_4 \rightarrow 0^+_1$	&	2403.05(22)	&	0.0029(3)	&	0.222(20)	&	0.0025(11)	&		&		&		\\
2488.59(11)	&	$>$0.55 	&	$4^+_2 \rightarrow 4^+_1$	&	208.46(21)	&	3.96(8)		&	60.3(12)	&		&	2.71(8)	&	52(7)	&		\\
			&				&	$4^+_2 \rightarrow 2^+_2$	&	445.90(21)	&	6.56(14)	&	100.0(22)	&		&	5.76(17)	&	100(3)	&		\\
			&				&	$4^+_2 \rightarrow 2^+_1$	&	1259.11(22)	&	3.87(10)	&	59.2(16)	&	$<$ 2.5	&	3.99(12)	&	67(3)	&	$<$2.8	\\
2677.10(11)	&	$>$0.28 	&	$2^+_5 \rightarrow 2^+_1$	&	1447.51(21)	&	0.035(3)	&	87(6)	&	$<$ 3.9	&	0.047(5)	&	86(4)	&	$<$3.7	\\
			&				&	$2^+_5 \rightarrow 0^+_1$	&	2677.18(20)	&	0.0405(25)	&	100(6)	&	$<$ 0.25	&	0.039(5)	&	100(4)	&	$<$0.23	\\
2733.53(11)	&	0.5$^{+0.6}_{-0.2}$ 	&	$4^+_3 \rightarrow 4^+_1$	&	452.7(3)		&	0.098(11)	&	6.2(7)	&	100$^{+70}_{-50}$	&		&		&		\\
			&				&	$4^+_3 \rightarrow 2^+_2$	&	690.89(21)	&	0.0199(18)	&	1.26(12)	&	2.5$^{+1.6}_{-1.4}$	&		&		&		\\
			&				&	$4^+_3 \rightarrow 2^+_1$	&	1504.0(3)	&	1.58(4)		&	100(3)	&	4.0$^{+2.6}_{-2.2}$	&	1.65(5)	&	100	&	4(6)	\\
2878.4(3)	&				&	$(5^-)^a \rightarrow 4^+_2$	&	598.2(3)	&	0.067(6)	&	100	&		&	0.069(11)	&	100(3)	&		\\
2903.53(12)	&	0.077$^{+0.020}_{-0.013}$ 	&	$2^+_6 \rightarrow 2^+_1$	&	1673.76(28)	&	0.0216(17)	&	67(7)	&	0.9(3)	&		&	37(2)	&	2.2(6)	\\
			&				&	$2^+_6 \rightarrow 0^+_1$	&	2903.46(21)	&	0.0324(24)	&	100(7)	&	0.7(2)	&	0.028(9)$^b$	&	100(4)	&	0.8(2)	\\
2963.04(12)	&				&	$4^+_4 \rightarrow 4^+_3$	&	229.6(4)	&	1.06(5)		&	1.80(8)	&		&	0.783(24)	&	1.38(4)	&		\\
			&				&	$4^+_4 \rightarrow 2^+_5$	&	286.02(20)	&	0.0371(17)	&	0.063(3)	&		&		&		&		\\
			&				&	$4^+_4 \rightarrow 4^+_2$	&	474.5(3)	&	3.10(7)		&	5.26(16)	&		&	3.00(10)	&	5.30(18)	&		\\
			&				&	$4^+_4 \rightarrow 2^+_4$	&	560.04(20)	&	1.161(26)	&	1.97(4)	&		&	0.99(4)	&	1.75(7)	&		\\
			&				&	$4^+_4 \rightarrow 2^+_3$	&	635.2(3)	&	1.73(4)		&	2.94(7)	&		&	1.77(6)	&	3.13(11)	&		\\
			&				&	$4^+_4 \rightarrow 3^-_1$	&	638.4(3)	&	1.60(4)		&	2.72(6)	&		&	1.37(4)	&	2.42(7)	&		\\
			&				&	$4^+_4 \rightarrow 4^+_1$	&	682.94(20)	&	59.0(13)	&	100.0(23)	&		&	56.6(17)	&	100(3)	&		\\
			&				&	$4^+_4 \rightarrow 2^+_2$	&	920.4(3)	&	0.476(12)	&	0.807(20)	&		&	0.506(21)	&	0.89(4)	&		\\
			&				&	$4^+_4 \rightarrow 2^+_1$	&	1733.56(22)	&	0.446(13)	&	0.76(5)	&		&		&		&		\\
2999.12(18)	&		&	$6^+_1 \rightarrow 4^+_2$	&			510.88(21)  &	0.094(8) 	& 	100(8) && 0.13(2) & 100(17) & \\	
		&		&	$6^+_1 \rightarrow 4^+_1$	&			718.57(21)  &	0.090(7) 	&			96(7) && 0.076(13) & 60(3) & \\
3056.88(14)	&				&	$2^+_7 \rightarrow 0^+_1$	&	3056.90(22)	&	0.0108(11)	&	100	&	0.5(2)	&		&	100(4)	&	0.5(2)	\\
3374.10(11)	&				&	$4^+_5 \rightarrow 4^+_4$	&	411.16(22)	&	0.045(5)	&	8.2(9)	&		&	0.037(7)	&	4.6(9)	&		\\
			&				&	$4^+_5 \rightarrow 2^+_6$	&	470.66(22)	&	0.00084(6)	&	0.154(10)	&		&		&		&		\\
			&				&	$4^+_5 \rightarrow 4^+_3$	&	640.50(21)	&	0.0222(11)	&	4.05(20)	&		&		&		&		\\
			&				&	$4^+_5 \rightarrow 2^+_5$	&	696.87(22)	&	0.0067(12)	&	1.23(23)	&		&		&		&		\\
			&				&	$4^+_5 \rightarrow 4^+_2$	&	885.43(20)	&	0.259(7)	&	47.3(12)	&		&	0.264(20)	&	33(3)	&		\\
			&				&	$4^+_5 \rightarrow 2^+_4$	&	971.0(3)	&	0.139(6)	&	25.4(10)	&		&	0.35(6)	&	44(8)	&		\\
			&				&	$4^+_5 \rightarrow 2^+_3$	&	1046.31(21)	&	0.070(7)	&	12.8(13)	&		&		&		&		\\
			&				&	$4^+_5 \rightarrow 4^+_1$	&	1094.10(20)	&	0.548(23)	&	100(4)	&		&	0.805(20)	&	100(3)	&		\\
			&				&	$4^+_5 \rightarrow 2^+_2$	&	1331.3(3)	&	0.0127(18)	&	2.3(3)	&		&		&		&		\\
			&				&	$4^+_5 \rightarrow 2^+_1$	&	2144.64(21)	&	0.108(11)	&	19.7(20)	&		&	0.121(5)	&	15.0(6)	&		\\
3397.46(13)	&				&	$4^+_6 \rightarrow 4^+_3$	&	663.92(20)	&	0.0338(16)	&	43.0(20)	&		&		&		&		\\
			&				&	$4^+_6 \rightarrow 4^+_2$	&	908.72(20)	&	0.079(4)	&	100(5)	&		&		&		&		\\
			&				&	$4^+_6 \rightarrow 2^+_4$	&	994.18(22)	&	0.0166(11)	&	21.1(14)	&		&		&		&		\\
			&				&	$4^+_6 \rightarrow 2^+_3$	&	1070.06(23)	&	0.035(3)	&	45(4)	&		&		&		&		\\
			&				&	$4^+_6 \rightarrow 4^+_1$	&	1117.3(3)	&	0.043(4)	&	54(5)	&		&		&		&		\\
			&				&	$4^+_6 \rightarrow 2^+_1$	&	2168.3(4)	&	0.008(3)	&	10(4)	&		&		&		&		\\
3460.21(11)	&				&	$4^+_7 \rightarrow 2^+_7$	&	403.41(22)	&	0.0047(4)	&	1.20(10)	&		&		&		&		\\
			&				&	$4^+_7 \rightarrow 2^+_6$	&	556.54(21)	&	0.0064(4)	&	1.64(11)	&		&		&		&		\\
			&				&	$4^+_7 \rightarrow 4^+_3$	&	726.62(21)	&	0.019(1)	&	4.93(27)	&		&		&		&		\\
			&				&	$4^+_7 \rightarrow 2^+_5$	&	783.10(22)	&	0.024(4)	&	6.0(11)	&		&		&		&		\\
			&				&	$4^+_7 \rightarrow 4^+_2$	&	971.6(6)	&	0.39(3)	&	100(8)	&		&	0.32(7)	&	96(21)	&		\\
			&				&	$4^+_7 \rightarrow 2^+_4$	&	1057.15(22)	&	0.0148(10)	&	3.79(26)	&		&		&		&		\\
			&				&	$4^+_7 \rightarrow 2^+_3$	&	1132.42(20)	&	0.090(3)	&	23.1(8)	&		&	0.099(9)	&	30(3)	&		\\
			&				&	$4^+_7 \rightarrow 4^+_1$	&	1180.21(20)	&	0.159(9)	&	40.7(23)	&		&	0.163(10)	&	49(3)	&		\\
			&				&	$4^+_7 \rightarrow 2^+_2$	&	1417.57(21)	&	0.0240(15)	&	6.1(4)	&		&	0.027(5)	&	8(2)	&		\\
			&				&	$4^+_7 \rightarrow 2^+_1$	&	2230.7(9)	&	0.280(11)	&	71.4(28)	&		&	0.333(11)	&	100(3)	&		\\
3592.15(11)	&				&	$4^+_8 \rightarrow 2^+_7$	&	535.21(22)	&	0.0036(3)	&	1.37(11)	&		&		&		&		\\
			&				&	$4^+_8 \rightarrow 4^+_3$	&	858.52(21)	&	0.108(4)	&	41.2(16)	&		&	0.117(20)	&	42(7)	&		\\
			&				&	$4^+_8 \rightarrow 2^+_5$	&	915.20(22)	&	0.0034(6)	&	1.30(26)	&		&		&		&		\\
			&				&	$4^+_8 \rightarrow 4^+_2$	&	1103.37(21)	&	0.083(7)	&	31.5(27)	&		&		&		&		\\
			&				&	$4^+_8 \rightarrow 2^+_3$	&	1189.25(22)	&	0.0140(8)	&	5.31(29)	&		&		&		&		\\
			&				&	$4^+_8 \rightarrow 2^+_3$	&	1264.34(20)	&	0.051(6)	&	19.4(24)	&		&	&		&		\\
			&				&	$4^+_8 \rightarrow 4^+_1$	&	1312.1(6)	&	0.181(6)	&	68.7(23)	&		&	0.187(9)	&	67(3)	&		\\
			&				&	$4^+_8 \rightarrow 2^+_2$	&	1549.52(20)	&	0.264(8)	&	100.0(29)	&		&	0.281(12)	&	100(4)	&		\\
			&				&	$4^+_8 \rightarrow 2^+_1$	&	2362.79(21)	&	0.064(3)	&	24.1(12)	&		&	0.068(4)	&	24(1)	&		\\
3704.34(11)	&				&	$4^+_9 \rightarrow 2^+_6$	&	800.76(20)	&	0.0168(9)	&	12.3(7)	&		&		&		&		\\
			&				&	$4^+_9 \rightarrow 4^+_3$	&	970.80(21)	&	0.110(5)	&	81(3)	&		&		&		&		\\
			&				&	$4^+_9 \rightarrow 2^+_5$	&	1027.16(22)	&	0.0043(8)	&	3.2(6)	&		&		&		&		\\
			&				&	$4^+_9 \rightarrow 4^+_2$	&	1215.71(21)	&	0.030(3)	&	22.4(23)	&		&		&		&		\\
			&				&	$4^+_9 \rightarrow 2^+_4$	&	1301.37(20)	&	0.0411(18)	&	30.3(14)	&		&	0.056(6)	&	37(4)	&		\\
			&				&	$4^+_9 \rightarrow 2^+_3$	&	1376.65(20)	&	0.0265(24)	&	19.5(18)	&		&	0.038(5)	&	25(3)	&		\\
			&				&	$4^+_9 \rightarrow 4^+_1$	&	1424.1(3)	&	0.019(4)	&	27(6)	&		&	0.021(5)	&	14(3)	&		\\
			&				&	$4^+_9 \rightarrow 2^+_2$	&	1661.57(21)	&	0.0292(24)	&	21.5(18)	&		&	0.041(6)	&	27(4)	&		\\
			&				&	$4^+_9 \rightarrow 2^+_1$	&	2475.06(20)	&	0.136(7)	&	100(5)	&		&	0.150(7)	&	100(5)	&		\\
3753.74(14)	&				&	$4^+_{10} \rightarrow 4^+_3$	&	1020.14(20)	&	0.070(3)	&	39.0(17)	&		&		&		&		\\
			&				&	$4^+_{10} \rightarrow 4^+_2$	&	1265.14(20)	&	0.064(6)	&	36(3)	&		&	0.138(9)	&	78(5)	&		\\
			&				&	$4^+_{10} \rightarrow 4^+_1$	&	1473.55(21)	&	0.179(6)	&	100(3)	&		&	0.177(8)	&	100(5)	&		\\
			&				&	$4^+_{10} \rightarrow 2^+_2$	&	1711.16(22)	&	0.0178(16)	&	9.9(9)	&		&		&		&		\\
			&				&	$4^+_{10} \rightarrow 2^+_1$	&	2524.3(3)	&	0.0061(11)	&	3.4(6)	&		&		&		&		\\
3816.19(15)	&				&	$4^+_{11} \rightarrow 2^+_6$	&	912.6(3)	&	0.00150(18)	&	1.72(21)	&		&		&		&		\\
			&				&	$4^+_{11} \rightarrow 4^+_3$	&	1082.8(3)	&	0.0063(6)	&	7.2(7)	&		&		&		&		\\
			&				&	$4^+_{11} \rightarrow 4^+_2$	&	1327.66(22)	&	0.0190(22)	&	21.8(26)	&		&		&		&		\\
			&				&	$4^+_{11} \rightarrow 4^+_1$	&	1536.1(4)	&	0.0069(12)	&	7.9(14)	&		&		&		&		\\
			&				&	$4^+_{11} \rightarrow 2^+_1$	&	2586.57(21)	&	0.087(5)	&	100(6)	&		&	0.096(6)	&	100	&		\\
3838.33(16)	&				&	$4^+_{12} \rightarrow 2^+_6$	&	934.7(3)	&	0.00085(15)	&	1.06(20)	&		&		&		&		\\
			&				&	$4^+_{12} \rightarrow 4^+_3$	&	1104.5(3)	&	0.0030(12)	&	3.8(15)	&		&		&		&		\\
			&				&	$4^+_{12} \rightarrow 4^+_2$	&	1350.2(4)	&	0.0081(16)	&	10.2(21)	&		&		&		&		\\
			&				&	$4^+_{12} \rightarrow 4^+_1$	&	1558.0(3)	&	0.0144(24)	&	18(3)	&		&		&		&		\\
			&				&	$4^+_{12} \rightarrow 2^+_1$	&	2608.96(21)	&	0.080(4)	&	100(5)	&		&	0.086(6)	&	100	&		\\

\midrule
\bottomrule
\hline
\hline
\multicolumn{10}{@{\extracolsep{\fill}}l}{\footnotesize{$^a$Listed as (4,5,6$^+$) in \cite{nndc} but suggested to be $5^-$ in Ref.~\cite{mikhailov, guaz_pt}. A 5$^-$ assignment also fits with the log$ft$ value in Table \ref{logft}.}} \\
\multicolumn{10}{@{\extracolsep{\fill}}l}{\footnotesize{$^b$Observed by Raman \textit{et al.} but not placed in level scheme.}}\\
\label{levels}
\end{longtable*}

\begin{table}[ht]
\caption{$\beta$-feeding intensity was measured to calculate log$ft$ values~\cite{logft}. Conversion electron coefficients were taken into account and were calculated using BrIcc~\cite{bricc}. The 2$^+$ levels with non-zero beta intensities are discussed in Section~\ref{results}.}
\begin{tabular*}{\linewidth}{c @{\extracolsep{\fill}} c c c c c}
\hline 
\hline
5$^+ \rightarrow J^\pi$ & Energy & $I_{\beta^-}$\% & log$ft$ & $I_{\beta^-}$\% & log$ft$\\ 
 &$[\text{keV}]$ & \multicolumn{2}{c}{This work} & \multicolumn{2}{c}{Ref.~\cite{nndc}}\\
\midrule
\hline
4$^+$ & 2280.21  & 19.7(24) & 5.92(6) & 22(3) & 5.83(6)\\ 

4$^+$ & 2488.59 & 10.24(20) & 6.033(18) & 8.5(4) & 6.058(23)\\ 
 
4$^+$ & 2733.53 & 0.17(6) & 7.58(16) & 0.67(7)  & 6.90(4)\\ 
 
($5^-$) & 2878.4 & 0.065(6) & 7.9(5) & 0.066(11)  & 7.78(8)\\

$2^{+}$ & 2903.53 & 0.027(3) & 10.08(6) & -  & -\\  
 
4$^+$ & 2963.04 & 66.2(8) & 4.752(8) & 62(3)  & 4.712(24)\\ 

6$^+$ & 2999.12 & 0.178(10) & 7.28(4) & 0.198(24) & 7.17(6)\\ 

2$^+$ & 3056.88 & 0.0021(11) & 10.84(22) & - & -\\

4$^+$ & 3159.28 & - & - & 0.104(18) & 7.25(8)\\
 
4$^+$ & 3374.10 & 1.17(3) & 5.98(3)  &1.51(8) & 5.78(3) \\ 
 
4$^+$ & 3397.46 & 0.207(8) & 6.70(3)  & - & -\\ 
 
4$^+$ & 3460.21 & 0.98(3) & 5.93(3) & 0.90(8) & 5.87(5) \\ 
 
4$^+$ & 3592.15 & 0.733(14) & 5.83(4) & 0.63(4) & 5.79(4)\\ 
4$^+$ & 3704.34 & 0.398(10) & 5.89(4) & 0.294(16) & 5.89(3)\\ 
4$^+$ & 3753.74 & 0.325(9) & 5.88(4) & 0.302(15) & 5.77(3)\\ 
4$^+$ & 3816.19 & 0.117(6) & 6.19(6) & 0.092(7) & 6.14(4)\\ 
$4^{+}$ & 3838.33 & 0.102(5) & 6.19(5) & 0.083(7) & 6.13(5)\\ 
\hline 
\hline
\end{tabular*}
\label{logft}
\end{table}

\subsection{285 keV triplet}

A triplet of 284.5-keV, 285.3-keV and 286.0-keV transitions was resolved using $\gamma$-$\gamma$ coincidences and gating on transitions in their respective cascades as shown in Figure~\ref{partial}. In the study by Raman \textit{et al.}~\cite{raman}, $\gamma$-$\gamma$ coincidences were not made and the 285.2~keV $\gamma$ ray which they observed was placed to the $2^+_3$, 2328-keV level. In another study by Mikhailov \textit{et al.}~\cite{mikhailov}, $\gamma$-$\gamma$ coincidences were measured and a 284.66-keV transition was placed to the $2^+_2$, 2043-keV level. They did not observe the 285.2-keV transition, despite populating the $2^+_3$, 2328-keV level. We observed 284.5-keV and 285.3-keV $\gamma$ rays directly in coincidence with each other making them difficult to separate. 
%Check this part again - Mark suggests reordering to give B(E2) after separation
The 284.5-keV transition was determined by gating from above on 446~keV and the relative branching ratio of 284.5~keV was determined, from the same gate, to be 1.33(6) compared to the previously adopted 2.5(2)~\cite{nndc}. This reduces the $B(E2; 2^+_2 \rightarrow 0^+_2$) from 39(7) W.u. to 21(4) W.u.,\ indicating a less collective transition than the previous value suggests.

The Compton edge of the 446-keV full-energy peak, occuring at $\sim$284~keV, made it difficult to isolate the 285.3-keV transition when gating from below on 2043~keV and on 813~keV. A gate was instead placed from above on 635~keV to isolate the 285.3-keV transition. The small contribution from the 284.4~keV transition in coincidence with 285.3~keV was taken into account.
%Although the 285.3-keV $\gamma$-ray feeds the 2043-keV level,  it was assumed that there was negligible contribution of 284.5~keV (due to the small branching ratio of 1.33~\%) in the 285.3-keV full-energy peak when gating on the 635-keV transition. 
It was assumed there was no $\beta$-feeding to the $2^+_2$ and $2^+_3$, based on the $\beta$-decay selection rules, such that the relative intensities could be determined from gating from above. The relative intensity of 285.3~keV was determined to be 0.038(14), half of the previous value of 0.081(10)~\cite{raman}.

\begin{figure}[ht]

\includegraphics[width=\columnwidth]{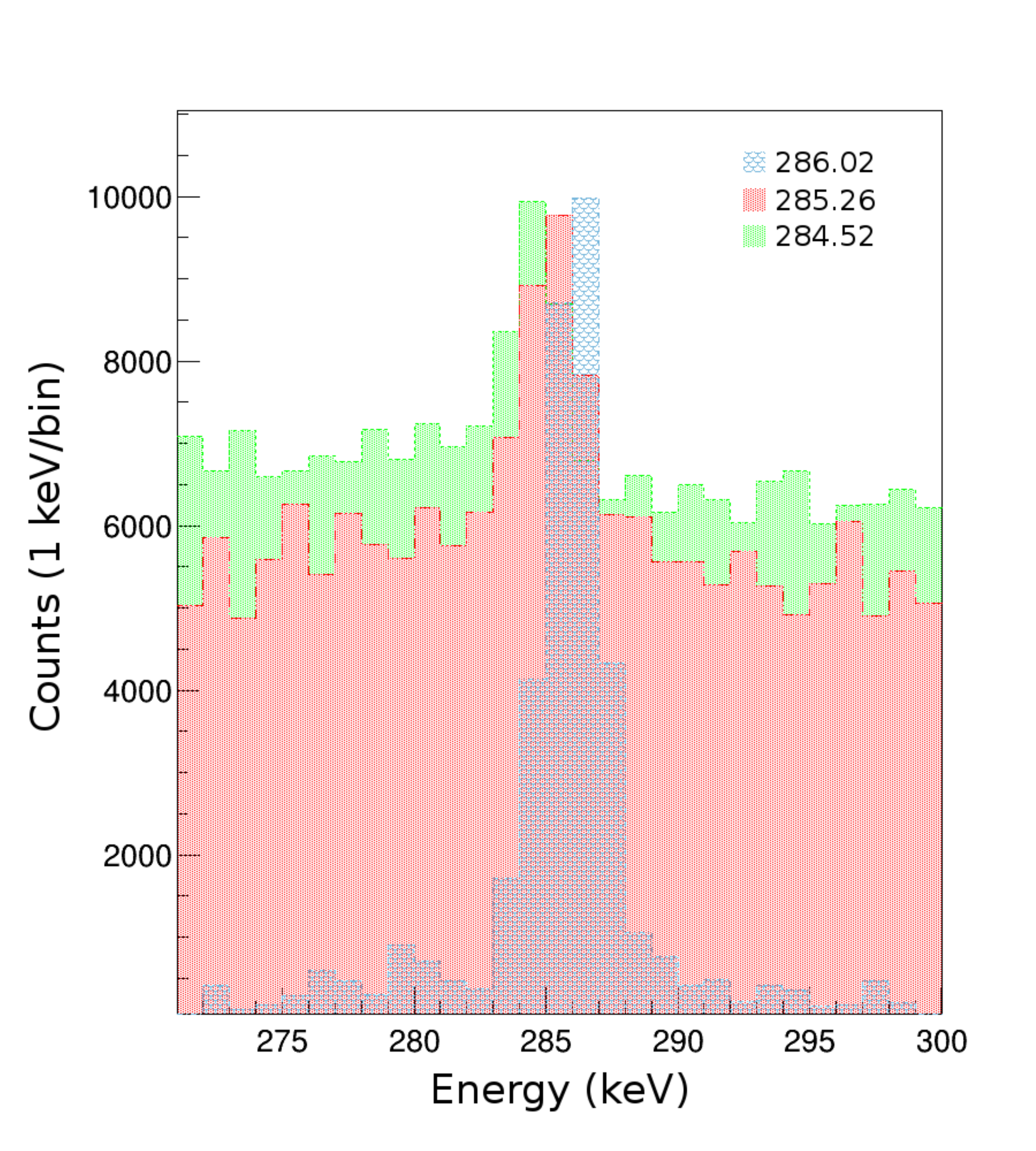}
\caption{Overlap of the transitions making up the 285-keV triplet. The individual spectra have been scaled to better show the energy differences. The corresponding coincidental transitions gated on to observe these energy shifts can be seen in Figure~\ref{partial} (represented by color).}
\label{triplet}
\end{figure}

The newly observed 286.0~keV $\gamma$-ray transition from the 2963.141~keV level is in a separate cascade and was far easier to decouple from the other two. A gate was placed below on the 2677.18~keV transition to isolate the 286.0~keV transition. Figure~\ref{triplet} shows an overlap of all three transitions and the subtle shifts of the peak centroids.

\subsection{Gamma-gamma angular correlations}
\label{sec_angcorr}

The geometry of GRIFFIN allows $\gamma$-$\gamma$ angular correlations to be performed, using 51 correlation angles between detector pairs. The statistics for $\gamma$-$\gamma$ angular correlations were not sufficient to assign spins to the levels with no definite spin assignment. 
 
Angular correlations for cascades with sufficient statistics were made to obtain $\delta$ values. Table~\ref{angcorr} summarizes the $\chi^2$ of the angular correlation fit and the $\delta$ which was determined through $\chi^2$ minimization. The form of the angular correlation fit is 
\begin{equation}
\label{eq1}
W(\theta)= A_{00}[1 + a_2P_2(cos\theta) + a_4P_4(cos\theta)]
\end{equation}
where $a_2$ and $a_4$ are coefficients which depend on the $\gamma$-ray multipolarities and their mixing ratios, $\theta$ is the angle between the successive $\gamma$ rays in a cascade, and $P_i(cos\theta)$ are Legendre polynomials.

\begin{figure}
     \includegraphics[width=\linewidth]{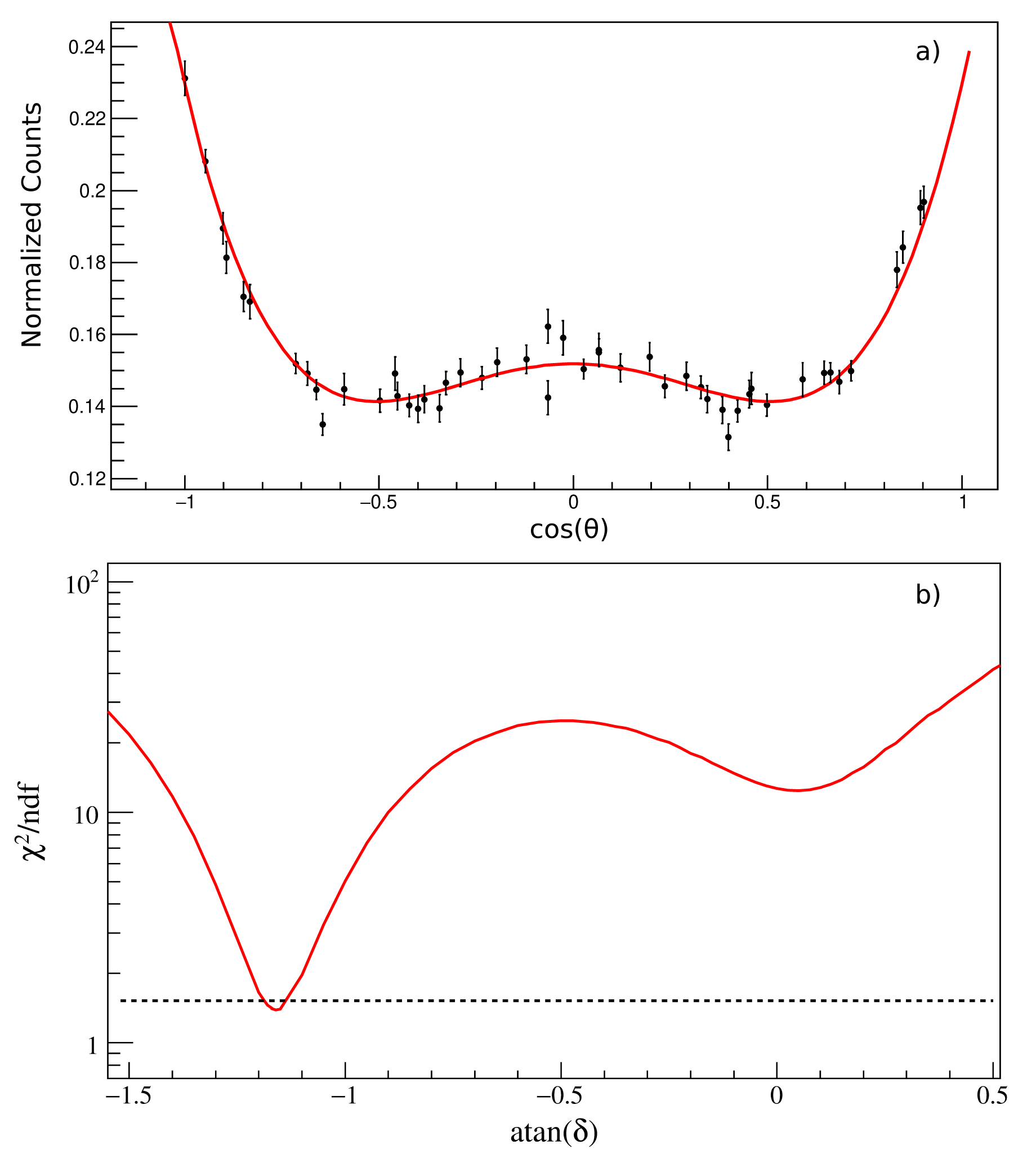}    
		\caption{Angular correlation (a) of the $2^+_2 \rightarrow 2^+_1 \rightarrow 0^+_1$ (813~keV-1230~keV coincidence) and its corresponding $\chi^2$ minimization plot (b) to determine the mixing ratio. In (b), the dashed line respresents the 3$\sigma$ limit to identify the mixing ratio for a given spin assignment. In this case, the $2^+_2 \rightarrow 2^+_1$ has a $\delta = -2.28(7)$ which agrees with the literature value of -2.34(16).} 
\label{angcorr813}
\end{figure}

Due to the finite size of the GRIFFIN detectors, the $a_2$ and $a_4$ coefficients are attenuated and energy-dependent attenuation factors need to be applied when fitting the angular correlations with Eq.~\ref{eq1}. 
Attenuation factors which have previously been determined in Method 4 of Ref.~\cite{jksmith_angcorr} were applied to these coefficients. However, these attenuation factors were obtained with a setup which did not include the Delrin shield. Therefore, simulations have been performed for several energy pairs with the inclusion of the Delrin shield. It was determined that the attenuation factors differentiated by at most $1\%$ which is within the uncertainty of our angular correlation fits. An example of the angular correlation for the $2_2^+\rightarrow 2_1^+ \rightarrow 0^+_1$ cascade is shown with the corresponding $\chi^2$ minimization in Figure~\ref{angcorr813}. .

The largest discrepancy which could impact future calculations, such as $B(E2)$ values, is the mixing ratio obtained for the 683-keV transition. The newly measured value of 0.224(3) would increase the $B(E2)$ by nearly a factor of six from the previous value of 0.09(5). In addition, the sign of the $\delta$ of the 1098-1230-keV transition differs from the one previously reported~\cite{nndc}.

\begin{table}

\caption{The angular correlations made in this study are summarized by the transition gated on, $E_{\gamma1} $, and the corresponding transition, $E_{\gamma2} $, used to make the angular correlation (eg. Figure~\ref{angcorr813}a).}
\begin{tabular}{l l l l l | l}
\hline
\hline 
$E_{\gamma1} $ & $E_{\gamma2} $ & $J_2\rightarrow J_1 \rightarrow J_0$ & $\chi^2_{\nu}$ & $\delta$ & $\delta$~\cite{nndc} \\ 
$[\text{keV}]$& $[\text{keV}]$ & & & &\\
\hline
1230 & 813 & $2_2^+\rightarrow 2_1^+\rightarrow 0_1^+$ & 1.39  & -2.28(7) & -2.34(16) \\
  & 1098 & $2_3^+\rightarrow 2_1^+\rightarrow 0_1^+$ & 0.86 & -14(4) & 56(31)\footnote{Listed as $1/\delta = 0.018(10)$~\cite{nndc}} \\
   & 1173 & $2_4^+\rightarrow 2_1^+\rightarrow 0_1^+$ & 1.21 & 0.85(3) & 1.07(9) \\
1050 & 208 & $4_2^+\rightarrow 4_1^+\rightarrow 2_1^+$ & 1.36 & -0.19(4) & -0.17(4) \\ 
 & 683 & $4_4^+\rightarrow 4_1^+\rightarrow 2_1^+$ & 1.15 & 0.224(3) & 0.09(5) \\  
\hline
\hline 

\end{tabular}
\label{angcorr}
\end{table}

\section{$sd$ IBM-2 calculations with mixing}

\label{sec:IBM}

As in previous studies on the mixing between the normal and intruder configuration in $^{114}$Sn\,\citep{spieker2018} and $^{116}$Sn\,\cite{petrache}, an $sd$ IBM-2 calculation with mixing was performed using the computer code \textsc{NPBOS}\,\citep{npbos}. 
Following the adopted approach for $^{114,116}$Sn\,\cite{spieker2018, petrache}, the intruder states in $^{118}$Sn are described using the corresponding Pd isotone, $^{114}$Pd, and the parameters determined in Ref.~\cite{pdibm}.
%Following the adopted approach for $^{114,116}$Sn\,\cite{spieker2018, petrache}, the corresponding Pd isotone, $^{114}$Pd, and the parameters determined in \cite{pdibm} to describe the intruder states in $^{118}$Sn. 
The $c_{4}^{\rho}$ parameter of the residual nucleon-nucleon interaction in the Hamiltonian was slightly adjusted from 0.10\,\cite{pdibm} to 0.05 to bring the intruder $4^+_2$ state closer in energy to the normal $4^+_1$ state. The $sd$ IBM-2 Hamiltonian $H$ has been described in detail in Ref.~\cite{pdibm, ibmhamiltonian} and the mixing Hamiltonian $H_{mix}$ in, $e.g.$, Ref~\cite{Del93a}. For completeness, they are presented here.

\begin{equation}
H = \varepsilon_{\pi}n_{d_{\pi}} + \varepsilon_{\nu}n_{d_{\nu}} + \kappa Q_{\pi} \cdot Q_{\nu} + M_{\pi \nu} + V_{\pi \pi} + V_{\nu \nu},
\end{equation}

where

\begin{equation}
Q_{\rho} = ( d^{\dagger} \times s + s^{\dagger} \times \tilde{d} )^{(2)}_{\rho} + \chi_{\rho} ( d^{\dagger} \times \tilde{d} )^{(2)}_{\rho},
\end{equation}

is the $sd$ IBM-2 quadrupole operator for $\rho =$ protons $\pi$ and neutrons $\nu$,

\begin{align}
M_{\pi \nu} = \frac{1}{2} \xi_2& \left[ \left( s_{\nu}^{\dagger} \times d^{\dagger}_{\pi} - d^{\dagger}_{\nu} \times s^{\dagger}_{\pi} \right)^{(2)}\notag \right. \\ 
& \left. \cdot \left( s_{\nu} \times \tilde{d}_{\pi} - \tilde{d}_{\nu} \times s_{\pi} \right)^{(2)} \right]^{(0)} \\
&+ \sum_{k=1,3} \xi_k \left[ \left( d^{\dagger}_{\nu} \times d^{\dagger}_{\pi} \right)^{(k)} \cdot \left( \tilde{d}_{\pi} \times \tilde{d}_{\nu} \right)^{(k)} \right]^{(0)} \notag
\end{align}

is the Majorana operator, and the residual nucleon-nucleon interaction is given by

\begin{align}
V_{\rho \rho} = &\sum_{L=0,2,4} \frac{1}{2} c_L^{\rho} \left[ \left( d_{\rho}^{\dagger} \times d_{\rho}^{\dagger} \right)^{(L)} \cdot \left( \tilde{d}_{\rho} \times \tilde{d}_{\rho} \right)^{(L)} \right]^{(0)} \nonumber
\\
&+ \frac{1}{2} v_0^{\rho} \left\{ \left[ \left( d^{\dagger}_{\rho}d^{\dagger}_{\rho}\right)^{(0)} \left( s_{\rho}s_{\rho}\right)^{(0)}\right]^{(0)}+h.c. \right\}  \\
&+ \sqrt{\frac{5}{2}} v_2^{\rho} \left\{ \left[ \left( d^{\dagger}_{\rho}d^{\dagger}_{\rho}\right)^{(2)} \left( \tilde{d}_{\rho}s_{\rho}\right)^{(2)}\right]^{(0)}+h.c. \right\} \nonumber
\\
&+\kappa_{\rho \rho} \left( Q_{\rho} \cdot Q_{\rho} \right)^{(0)}. \notag
\end{align}

To reduce the number of parameters, Kim {\it et al.} chose $c_L^{\pi} = c_L^{\nu} = c_L$ and $\kappa_{\rho \rho} = v_0^{\rho} = v_2^{\rho}= 0$ for the Pd isotopes\,\cite{pdibm}. For the normal states in $^{118}$Sn, we found the following parameters: $e_{\nu} = 1.26$\,MeV, $\kappa_{\nu \nu} = -0.0002$\,MeV, $\chi_{\nu} = -0.005$\,MeV, $c_0^{\nu} = -0.45$\,MeV, $c_2^{\nu} = 0$\,MeV, and $c_4^{\nu} = -0.11$\,MeV. All other parameters were set to 0. This parameter choice is, in fact, similar to the one used for $^{114,116}$Sn\,\cite{spieker2018, petrache}. The normal and intruder states were calculated separately and were then admixed using the mixing Hamiltonian $H_{mix}$ with $\alpha = 0.25$ and $\beta = 0$:

\begin{equation}
H_{mix} = \alpha \left( s_{\pi}^{\dagger} s_{\pi}^{\dagger} + s_{\pi} s_{\pi} \right)^{(0)} + \beta \left( d_{\pi}^{\dagger} d_{\pi}^{\dagger} + \tilde{d}_{\pi} \tilde{d}_{\pi} \right)^{(0)}.
\end{equation}

The energy gap $\Delta$ between the two configurations was set to 2.45\,MeV, which is the same value as the one previously reported for $^{116}$Sn\,\cite{petrache}.

After mixing, $E2$ transitions between the excited states were calculated using the $T(E2)$ operator following the consistent-$Q$ formalism:

\begin{equation}
\label{eq:e2}
T(E2) = e_0 \left( e_{\pi_0} Q_{\pi} + e_{\nu_0} Q_{\nu}\right) + e_2 \left( e_{\pi_2} Q_{\pi} + e_{\nu_2} Q_{\nu} \right) .
\end{equation}

These parameters were used to obtain the results discussed in the following section: $e_0 = 1$, $e_{\nu_0} = 0.07\,\mathrm{eb^2}$, $e_{\pi_0} = 0\,\mathrm{eb^2}$, $e_2 = 1.43$, $e_{\nu_2} = 0.10\,\mathrm{eb^2}$, and $e_{\pi_2} = 0.042\,\mathrm{eb^2}$.

\section{Discussion of the results}

Recent, detailed experimental studies by Garrett {\it et al.}\,\cite{Gar19a} showed that, in addition to the often discussed proton 2p-2h structure, multiple structures with proton multiparticle-multihole character coexist at comparably low excitation energies in $^{110,112}$Cd ($Z = 48$). The experimental results were supported by Beyond-Mean-Field (BMF) calculations, which predicted 2p-2h character for both deformed band structures built on the $0^+_2$ and $0^+_3$ states in $^{112}$Cd ($N = 64$). In the previous IBM studies\,\cite{spieker2018, petrache}, it was assumed that the intruder band structure in $^{114,116}$Sn ($N=64,66$) was caused by proton 2p-2h excitations across the $Z = 50$ shell closure and that it should resemble the yrast structure of the 0p-4h Pd isotopes. As pointed out by Garrett {\it et al.}, their results also suggest that the ground-state structure of Pd nuclei is more complex than the simple, spherical $\pi$(4h) configuration\,\cite{Gar19a}. This cannot be accounted for in coventional $sd$ IBM-2 calculations with mixing. However, as pointed out by Kim {\it et al.}\,\cite{pdibm}, the more neutron-rich Pd isotopes ($N \geq 64$) are clearly non-vibrational nuclei and closer to the $\gamma$-soft, O(6), limit of the IBM. It is worth noting that the recent measurement of the lifetimes of the yrast 4$^+$ and 6$^+$ states in $^{114}$Pd~\cite{114pd_newtime} are consistent with a rigid triaxial structure.

\begin{figure}
\centering
\includegraphics[width=\linewidth]{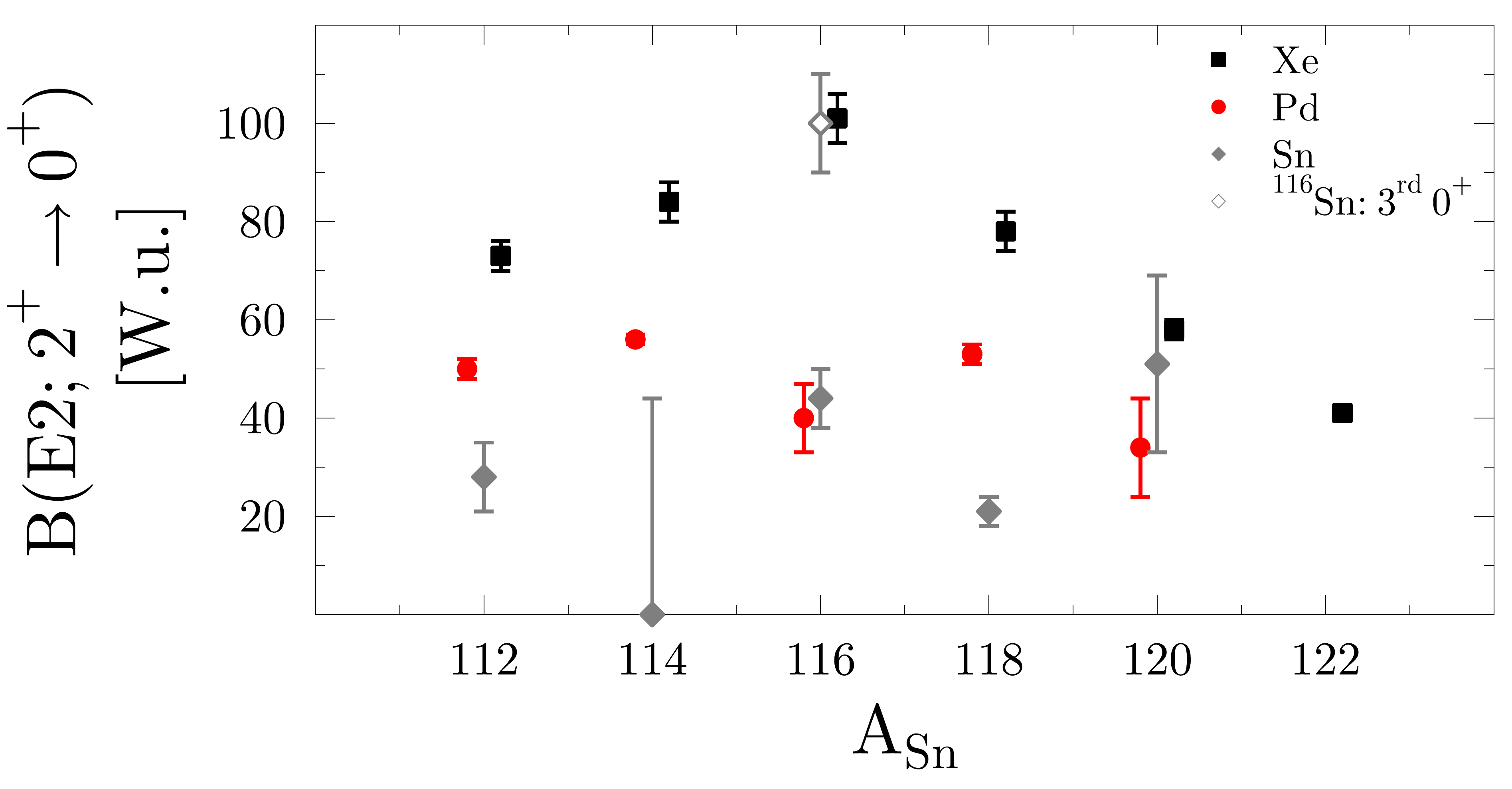}
\caption{\label{fig:be2_intruder}{$B(E2;2^+ \rightarrow 0^+)$ reduced transition strength between yrast states in Pd (red circles) and Xe (black squares), and between the $2^+_{\mathrm{\mathrm{intr.}}}$ and $0^+_i$ states in Sn ($i = 2$ (filled diamonds)). The $B(E2)$ values for all Xe and Pd isotopes are from~\cite{b(e2)comp} except $^{114}$Pd which is from ~\cite{114pd_newtime}. The values for $^{112}$Sn and $^{114}$Sn (upper limit) are from~\cite{spieker2018}, $^{116}$Sn from~\cite{pore116}, and $^{120}$Sn from data in the ENSDF~\cite{nndc120sn}. The $B(E2;2^+_{\mathrm{intr.}} \rightarrow 0^+_3)$ value (open diamond) is currently only known in $^{116}$Sn\,\cite{pore116, petrache}.}} 
 \end{figure}

\begin{figure}[ht]
\centering
\includegraphics[width=\linewidth]{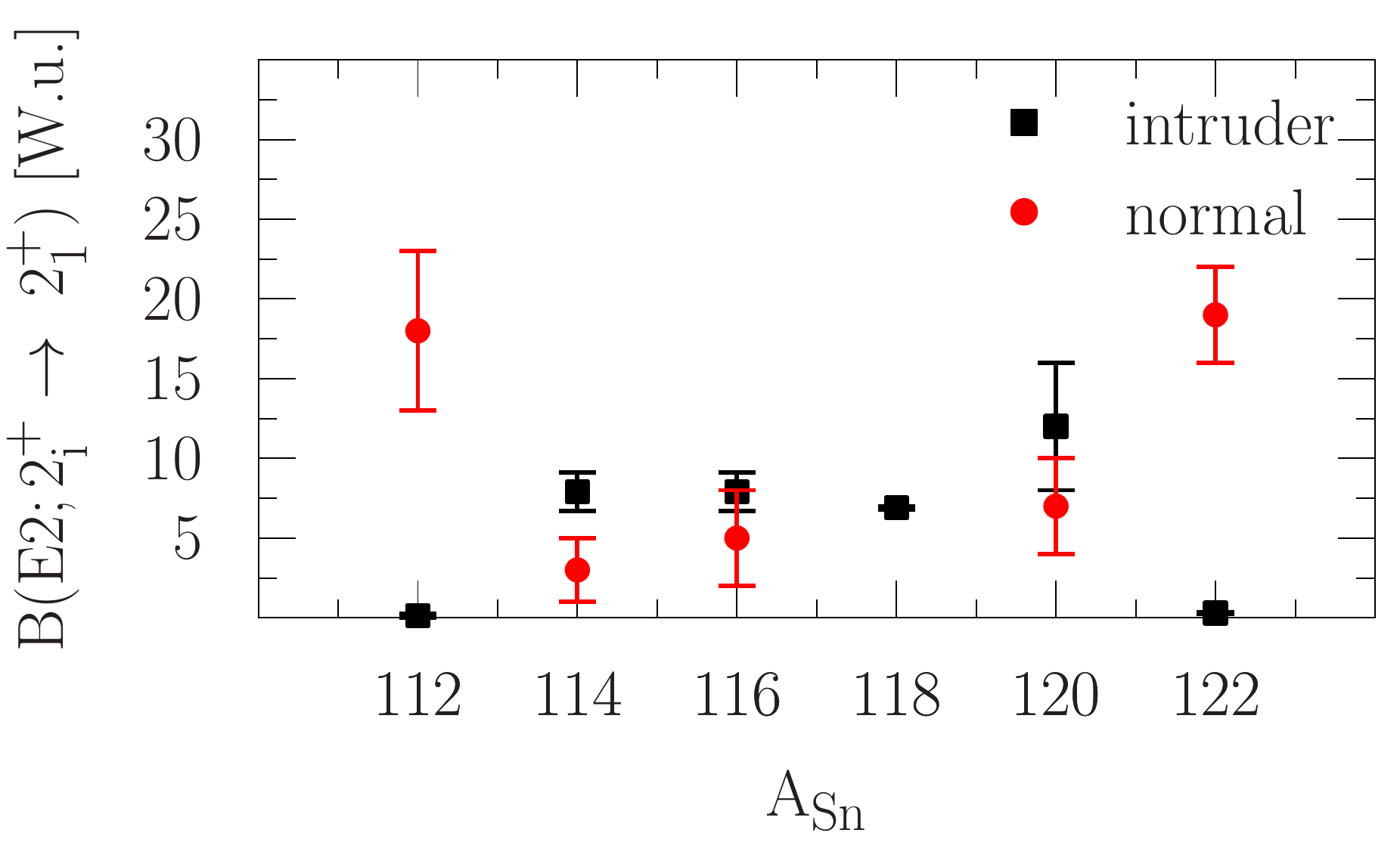}
\caption{\label{fig:be2_fragmentation}{$B(E2;2^+_i \rightarrow 2^+_1)$ reduced transition strength in the even-A Sn isotopes~\cite{spieker2018,pore116,nndc120sn,nndc122sn} to illustrate how strength originally attributed to the normal structure gets fragmented between the low-lying $2^+$ states once the $2^+_{\mathrm{\mathrm{intr.}}}$ drops below the $2^+_{\mathrm{norm.}}$ in $^{114}$Sn. In $^{122}$Sn, the intruder structure moves up energetically and the strength becomes concentrated in one state again.}} 
 \end{figure}

The experimental $B(E2;2^+ \rightarrow 0^+)$ reduced transition strengths for the yrast $2^+$ to the ground state in Pd (``$\pi$(4h)'') and Xe (``$\pi$(4p)'') as well as from the intruder $2^+_{\mathrm{\mathrm{intr.}}}$ to the $0^+_2$ state in Sn are compared in Fig.\,\ref{fig:be2_intruder}. In terms of absolute magnitude, the experimental $B(E2;2^+_{\mathrm{\mathrm{intr.}}} \rightarrow 0^+_2)$ strength is indeed more comparable to the strength in the Pd isotopes than the one in the Xe isotopes. 
It is likely the halflife for the yrast 2$^+$ state in $^{112}$Pd is too long\,\citep{114pd_newtime};  a shorter value would yield a smoother trend as observed in the Xe isotopes.
The decay of the $2^+_{\mathrm{intr.}}$ to the $0^+_3$ state has only been observed in $^{116}$Sn. From this observation, the $0^+_3$ was assigned as the bandhead of the intruder structure in $^{116}$Sn\,\cite{pore116}. Previous IBM studies supported this interpretation\,\cite{petrache} and pointed at a possibly similar scenario in $^{114}$Sn\,\cite{spieker2018}. The $0^+_2$ and $0^+_3$ are, however, strongly mixed. This is also why they share the $B(E2;0^+_i \rightarrow 2^+_1)$ transition strength in the IBM. Otherwise, the intruder state would show no collective transition strength to the $2^+_1$ state. Unlike for the $2^+_{\mathrm{\mathrm{intr.}}}$ and the second excited $2^+$ state of the normal configuration, the $\Delta L = 0$ and $\nu = \tau$ selection rules do not forbid mixing between the states\,\cite{Jol95a, Leh95a}. In $^{116}$Sn, the $B(E2;2^+_{\mathrm{\mathrm{intr.}}} \rightarrow 0^+_3) = 100(8)$\,W.u. matches the one observed in the yrast structure of $^{120}$Xe (see Fig.\,\ref{fig:be2_intruder}). The experimentally observed $B(E2;4^+_i \rightarrow 2^+_{\mathrm{\mathrm{intr.}}})$ strengths in $^{116}$Sn are, however, significantly lower than the strength predicted by the IBM and are only comparable to the yrast value in $^{120}$Xe when summed. At the same time, the summed $B(E2;2^+_{\mathrm{\mathrm{intr.}}} \rightarrow 0^+_i)$ value ($i = 2,3$) would significantly exceed the expectations. As pointed out in the previous studies~\cite{petrache,spieker2018}, strong mixing between the two configurations is observed and much more data are needed to draw more definite conclusions. 
Since the $B(E2)$ transition strength is shared between mixed states (compare Fig.~\ref{fig:be2_fragmentation}), if only part of the transition strength is observed experimentally, clear structure assignments can be challenging.
 
 \begin{table}[ht]
\caption{$sd$ IBM-2 calculations with mixing for low-lying excited states in $^{118}$Sn compared to experimental data. The lifetimes were taken from the ENSDF~\cite{nndc} as were mixing ratios $\delta$ for any transition which we were not able to obtain a $\delta$ value for. States predicted to originate from the intruder structure are marked with $\#$, states from the normal configuration with $\dagger$. Note, however, that most of the states are strongly mixed and often contain comparable amplitudes of the normal and intruder configuration in their wavefunction.}
\begin{tabular*}{\columnwidth}{r @{\extracolsep{\fill}} l l l l}
\hline 
\hline
$J_i^{\pi} \rightarrow J^{\pi}_f$ & $E_x$ & $E_{x,IBM}$ & $B(E2)$ & $B(E2)_{IBM}$ \\ 
 
& $[\text{keV}]$ & $[\text{keV}]$ & $[\text{W.u.}]$ & $[\text{W.u.}]$  \\ 
\hline 
$2^+_1 \rightarrow 0_1^+$ & 1230 & 1231$^{\dagger}$ & 12.1(5)  & 10 \\ 
$0^+_2 \rightarrow 2_1^+$ & 1758 & 1714$^{\#}$ & 19(3) & 16 \\ 
$2^+_2 \rightarrow 0_1^+$ & 2043 & 2098$^{\#}$ & 0.072(10) & 0.008 \\
$ \rightarrow 2_1^+$ &  &  & 7.2(10) & 0.13 \\ 
$ \rightarrow 0_2^+$ &  &  & 21(4) & 35 \\ 
$0^+_3 \rightarrow 2_1^+$ & 2057 & 2099$^{\dagger}$ &  & 10 \\
$ \rightarrow 2_2^+$ &  &  & & 11 \\ 
$4^+_1 \rightarrow 2_1^+$ & 2280 & 2270$^{\dagger}$ & 17(3) & 19 \\ 
$ \rightarrow 2_2^+$ &  &  & 16(3) & 18 \\ 
$2_3^+ \rightarrow 0_1^+$ & 2328 & 2256$^{\#}$ & $<$ 0.19 & 0.00010 \\ 
$ \rightarrow 2_1^+$ &  &  & $<$ 42 & 16 \\ 
$ \rightarrow 0_2^+$ &  &  & $<$ 26 & 0.5 \\ 
$ \rightarrow 2_2^+$ &  &  & $<$ 762 & 29 \\ 
$2^+_4 \rightarrow 0_1^+$ & 2403 & 2728$^{\dagger}$ &0.0025(11) & 0.0002 \\
$ \rightarrow 2_1^+$ &  &  & 17(7) & 4\\
$ \rightarrow 0_2^+$ &  &  & 12(5) & 0.05\\
$ \rightarrow 2_2^+$ &  &  & $<$ 81 & 26 \\
$ \rightarrow 2_3^+$ &  &  &  & 0.6 \\
$4^+_2 \rightarrow 2_1^+$ & 2489 & 2702$^{\#}$ & $<$ 2.5 & 1.2 \\
$ \rightarrow 2_2^+$ &  &  & $<$ 764 & 39 \\
$ \rightarrow 4_1^+$ &  &  & $<$ 673 & 0.3 \\
$4^+_3 \rightarrow 2_1^+$ & 2733 & 2903$^{\#}$ & 4$^{+5}_{-4}$ & 0.014 \\
$ \rightarrow 2_2^+$ &  &  & 3(3) & 0.09\\
$ \rightarrow 4_1^+$ &  &  & 100$^{+120}_{-100}$ & 19\\
$ \rightarrow 2_3^+$ &  &  &  & 28\\
$ \rightarrow 4_2^+$ &  &  &  & 12 \\ 
$6^+_1 \rightarrow 4_1^+$ & 3000 & 3034$^{\#}$ & &42 \\
$ \rightarrow 4_2^+$ &  &  &  & 20\\

\hline 
\hline
\end{tabular*}
\label{calcs}
\end{table}

This work adds $^{118}$Sn to the study and discusses possible mixing between the normal and intruder configurations based on a comparison of the new experimental data and IBM calculations, which were introduced above. In Sn, the intruder structure is energetically lowest in $^{116,118}$Sn (compare\,\cite{petrache}). No microscopic calculations as for $^{110,112}$Cd are presently available. Large-scale Monte-Carlo-Shell-Model (MCSM) calculations did, however, highlight the importance of proton excitations in Sn to understand the evolution of the $B(E2;2^+_1 \rightarrow 0^+_1)$ transition probabilities and showed that a second-order phase transition appears to take place from a moderately deformed to a spherical ground state for the more neutron-rich Sn isotopes when passing $^{116}$Sn\,\cite{togashi2018}. The results of the $sd$ IBM-2 calculations with mixing and the comparison to the corresponding experimental data are presented in Table~\ref{calcs}. The parameters are given in Sec.\,\ref{sec:IBM}.

Despite a few discrepancies, whose origin was already discussed for $^{116}$Sn\,\cite{petrache} and attributed to the selection rules of the O(5) symmetry present in the U(5) and O(6) group chains of the IBM\,\cite{Jol95a, Leh95a}, the agreement between the IBM results and the data is good both for the energies and transition probabilities (see Table\,\ref{calcs}). In contrast to $^{116}$Sn, only one excited $0^+$ state is found below the $2^+$ state of the lowest-lying intruder structure. A significant decrease of the $B(E2;2^+_{\mathrm{\mathrm{intr.}}} \rightarrow 0^+_2)$ strength by about a factor of 2 as compared to $^{116}$Sn is also observed.
%The previous data~\cite{nndc} showed the same $B(E2;2^+_1 \rightarrow 0^+_1)$ strength decrease in the corresponding Pd isotope, however more recent lifetime measurements for the $2^+_{\mathrm{intr.}}$ state in $^{114}$Pd~\cite{114pd_time,114pd_newtime} suggests larger collectivity (compare Fig.\,\ref{fig:be2_intruder}). 
To account for this experimental observation, the effective proton charge has been set to $e_{\pi_2} = 0.042$\,eb$^2$ ($e_{\pi_2} = 0.161$\,eb$^2$ in $^{116}$Sn and 0.105\,eb$^2$ in $^{114}$Sn). All other parameters in the $E2$ operator (compare Eq.\,(\ref{eq:e2})) are the same as in the studies of $^{114,116}$Sn\,\cite{spieker2018, petrache}. It should be noted that $e_{\pi_2}$ was drastically changed in $^{116}$Sn to obtain two very collective $B(E2;2^+_{\mathrm{intr.}}\rightarrow 0^+_{2,3})$ values\,\cite{petrache}. It remains to be seen if a similar effect will be observed in $^{114}$Sn. Based on all presently available experimental data in $^{118}$Sn, a scenario, where due to mixing the $2^+_{\mathrm{intr.}}$ would drop below the $0^+_{\mathrm{intr.}}$, is not favoured. With support from the IBM calculation the $0^+_2$ state is identified as the bandhead of the intruder structure in $^{118}$Sn and, in doing so, can also account for the collective $B(E2;0^+_2 \rightarrow 2^+_1)$ (see Table\,\ref{calcs}) generated due to mixing between the normal and intruder $0^+$ states. 
%Furthermore, the evolution of the stable Sn's excitation energies of the other bandmembers reveals a global minimum for the intruder configurations lowest $2^+$ state in $^{118}$Sn.
Furthermore,  the evolution of the excitation energy of the intruder bands in stable Sn isotopes reveals a global minimum for the lowest 2$^+$ intruder state in $^{118}$Sn.
Compared to $^{116}$Sn, lower excitation energies for the $4^+$ and $6^+$ intruder states are also observed\,\cite{nndc}.
%From the evolution of the excitation energies of the other bandmembers it can actually be seen that the intruder configuration further drops in energy in $^{118}$Sn. For the stable Sn isotopes, the $2^+_{\mathrm{\mathrm{intr.}}}$ reaches its global minimum and also the $6^+_1$ is lower in energy than in $^{116}$Sn\,\cite{nndc}.
To further characterize the intruder structure and mixing with the normal configuration in $^{118}$Sn, lifetime measurements for the $0^+_3$, $2^+_3$, $4^+_2$, and $6^+_1$ states are crucial. Additionally, precise measurements of branching ratios and of multipole mixing ratios $\delta$ in coincidence experiments are needed.
% Experimental challenges were discussed in Sec.~\ref{results}. 
The revised and smaller $E2$ branching ratio for the $6^+_1$ state of 5.8(8) is now closer to the IBM result of 2.1 but still significantly larger.

\section{Conclusion}
The study of the $\beta$-decay of $^{118}$In to low-lying excited states in $^{118}$Sn was performed using the GRIFFIN spectrometer at the TRIUMF-ISAC facility. We were able to identify 99 $\gamma$-ray transitions from 23 levels, of which 43 transitions and one level are newly observed. Using GRIFFIN's full array of 16 HPGe clover detectors allowed us to resolve three transitions near 285~keV which led to the reduction of the intensities previously assigned to 284.5~keV and 285.3~keV by half. This ultimately reduced the $B(E2; 2^+_{\mathrm{intr.}} \rightarrow 0^+_{\mathrm{intr.}})$ value from 39(7)~W.u. to 21(4)~W.u.. Less collectivity in the $2^+_{\mathrm{intr.}}$ state is not entirely supported by the present calculations, although it is possible that the $B(E2)$ strength is fragmented between the $2^+_{\mathrm{intr.}}$ and $2^+_{3,4}$ states.

%The experimental results were compared to $sd$ IBM-2 calculations with mixing which compare nicely to the previous calculations on $^{114,116}$Sn.
The experimental results were in reasonably good agreement with the predictions of  $sd$ IBM-2
calculations with mixing, similar to previous comparisons for $^{114,116}$Sn. 
The $0^+_3$ state in $^{114,116}$Sn had previously been suggested to be the bandhead of the $\pi$ 2p-2h intruder band. However, this isn't the case in $^{118}$Sn. Despite strong mixing with the $0^+_3$ state, the $0^+_2$ is still considered the bandhead and is supported by the IBM. Given the strong configuration mixing of many of the other states, it is difficult to conclusively assign structure. 
%In future campaigns, lifetime measurements for the even-even stable Sn excited states would be advantageous to better understand the configuration mixing within and the structural evolution between these nuclei.
In future campaigns, lifetime measurements of excited states in  even-even stable Sn isotopes would provide a better understanding of configuration mixing and the evolution of the structure of these nuclei.

\section*{Acknowledgments}

The infrastructure of GRIFFIN has been funded through contributions from the Canada Foundation for Innovation, TRIUMF, University of Guelph, British Columbia Knowledge Development Fund and the Ontario Ministry of Research and Innovation. TRIUMF receives funding through a contribution agreement through the National Research Council Canada. This work was supported by the Natural Sciences and Engineering Research Council of Canada. Mark Spieker acknowledges support by the National Science Foundation under contract No. PHY-1565546 (NSCL).

\bibliography{Sn118_manuscript_KO} 
\bibliographystyle{apsrev}

\end{document}